# Neutronic Chain Reactions for Polonium-210 Production

Solomon Lim[1]

[1] *National University of Singapore High School of Mathematics and Science*

## Abstract

The production of the industrially significant radionuclide polonium-210 from the neutron irradiation of bismuth metal and the subsequent beta decay of bismuth-210 is highly inefficient due to the small neutron capture cross section of bismuth-209. In this paper, we report a previously undescribed self-sustaining nuclear chain reaction involving self-propagating neutron multiplication in bismuth salts that allow for rapid and cost-effective production of polonium-210. The reaction proceeds in a cycle of three alternating elementary steps – the capture of neutrons by bismuth-209 and the subsequent formation of polonium-210, the emission of high-energy alpha particles by polonium-210, and the production of more neutrons from (α,n) and (n,2n) reactions on light element and bismuth-209 nuclei respectively. Furthermore, the high hydrogen density of the compound also confers it intrinsic neutron moderation properties, increasing the neutron capture cross section of bismuth-209 at thermal neutron energies. The chain reaction was proven to have successfully occurred by irradiating a sample of the bismuth salt with a 80 μCi neutron source and monitoring the activity levels of the reaction. It was found that the activity of the reaction increased exponentially after an initial stable period following a derived formula for polonium production trends for the reaction, thus validating the occurrence of the reaction. Furthermore, alpha spectroscopy confirmed that polonium-210 had been produced by characterising the 5.30 MeV alpha emission peak of the reaction in addition to using beta spectroscopy to identify the parent nuclide bismuth-210, further proving that the reaction was successful. Hence, this paper reports the successful initiation and characterisation of a novel nuclear chain reaction, and its potential applications offered by a method of rapidly producing large quantities of polonium-210.

## 1. Introduction

In 1931, neutrons were first produced by exposing beryllium to alpha particles from polonium-210 [1, 28, 29]. Since then, this method of neutron generation by irradiating light elements such as beryllium, lithium, and fluorine with an intense alpha source has been incorporated into radioisotopic neutron sources used in nuclear laboratories throughout the world. The mechanism by which the neutrons are produced proceeds via an (α, n) reaction which requires a source of high energy alpha particles [2, 26, 27], such as the $^9$Be(α, n)$^{12}$C reaction in beryllium. In particular, polonium-210 has been frequently used as the alpha source for such applications due to its high alpha emission energy of 5.4 MeV [30, 31], as well as a high specific activity of 4 kCi/g, which stems from its relatively low atomic mass and short half-life of 138 days [3]. Despite its toxicity, its intense alpha radioactivity also makes it an ideal candidate for a wide range of applications, such as in neutron initiators in nuclear weapons [32], medical isotopes in radiotherapy [33], energy cells in radioisotopic thermoelectric generators [34], and ionising antistatic devices [35] (Fig. 1). All these applications mean that polonium-210 is an industrially significant radionuclide, and thus producing it in quantities that can accommodate global demand is of practical importance. However, polonium-210 can only be produced in microgram amounts in nuclear reactors via neutron irradiation of bismuth-209, which is slow due to the small neutron absorption cross section of the $^{209}$Bi(n, γ)$^{210}$Bi reaction, measured to be 20.5 ± 1.1 millibarns [4,5].

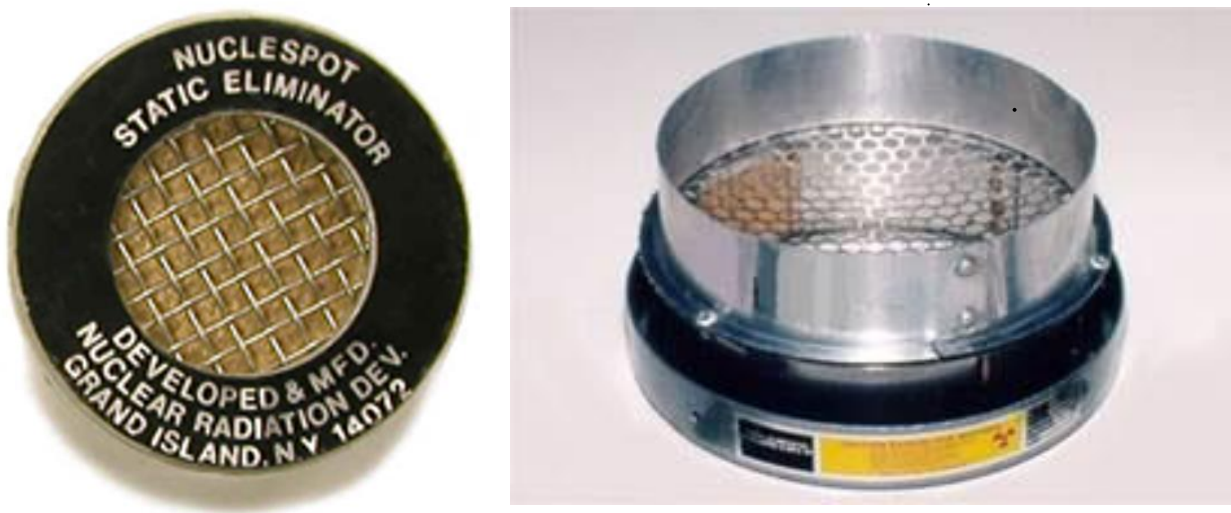

*Figure 1: Commercially available antistatic devices containing polonium-210*

Upon closer inspection of the aforementioned nuclides, a cyclic relationship is found: polonium-210 emits alpha particles that produce neutrons on impact with light elements, and neutrons can be used to produce polonium-210. Thus, a chain reaction can be initiated in a mixture of atoms of bismuth-209 and light elements, using an initial neutron source to transmute the bismuth atoms into that of polonium, which would themselves emit high-energy alpha particles that impact the light elements, causing further neutron production. Furthermore, $^{209}$Bi and $^{9}$Be are also able to cause neutron multiplication via a (n, 2n) reaction with fast neutrons [6], increasing the neutron population of the system. To be suitable for sustaining neutron multiplication, the bismuth salt should contain a light element with high neutron yields from (α, n) reactions [7], e.g. $^{27}$Al, $^{19}$F, $^{10}$B, $^{9}$Be, as well as a high hydrogen content for neutron moderation, as the $^{209}$Bi neutron capture cross section for $^{210}$Po formation is greatly increased at thermal neutron velocities. Hence, bismuth beryllium acetate was used due to its high beryllium content, which is ideal for alpha-to-neutron conversion, as well as its intrinsic neutron multiplication and moderation properties.

The validity of this chain reaction was investigated by exposing a sample of bismuth beryllium acetate to the neutron emissions from a weak neutron source (80 μCi) and measuring the activity levels of the reaction mixture. It is expected that in the case of a non-proliferating reaction, the activity of the target will approach an equilibrium before decreasing, while the successful propagation of the chain reaction would result in a sigmoidal increase of polonium-210 produced in the reaction mixture due to the autocatalytic nature of the chain reaction [36, 37]. It was found that the pattern of increase in radioactivity of the irradiated bismuth beryllium acetate samples were characteristic of a self-sustaining chain reaction as described by a derived equation (proof and formulation in Appendix 1). Furthermore, alpha and beta spectrometry was also used to further confirm the propagation of neutron multiplication through the bismuth salt sample by characterising the alpha emission spectrum of polonium-210 and the beta spectrum of bismuth-210.

## 2. Materials and Methods

2.1. Bismuth salt selection and synthesis

Selection of the bismuth salt to be used to sustain the chain reaction was achieved by considering the various properties that the salt would need to possess in order to be able to allow the chain reaction to propagate. Firstly, upon considering the ($\alpha$,n) reaction yields of the light elements from lithium to potassium [7] (Fig. 2), 3 elements were selected which have a substantially higher yield than the rest – beryllium, boron, and fluorine. The use of boron would be unfavourable due to its high neutron absorption cross section of 749 barns [16, 38], which would have absorbed the neutrons much faster than bismuth atoms and thus cause the chain reaction to fail. Between beryllium and fluorine, beryllium was chosen due to its markedly higher ($\alpha$,n) reaction yield [39].

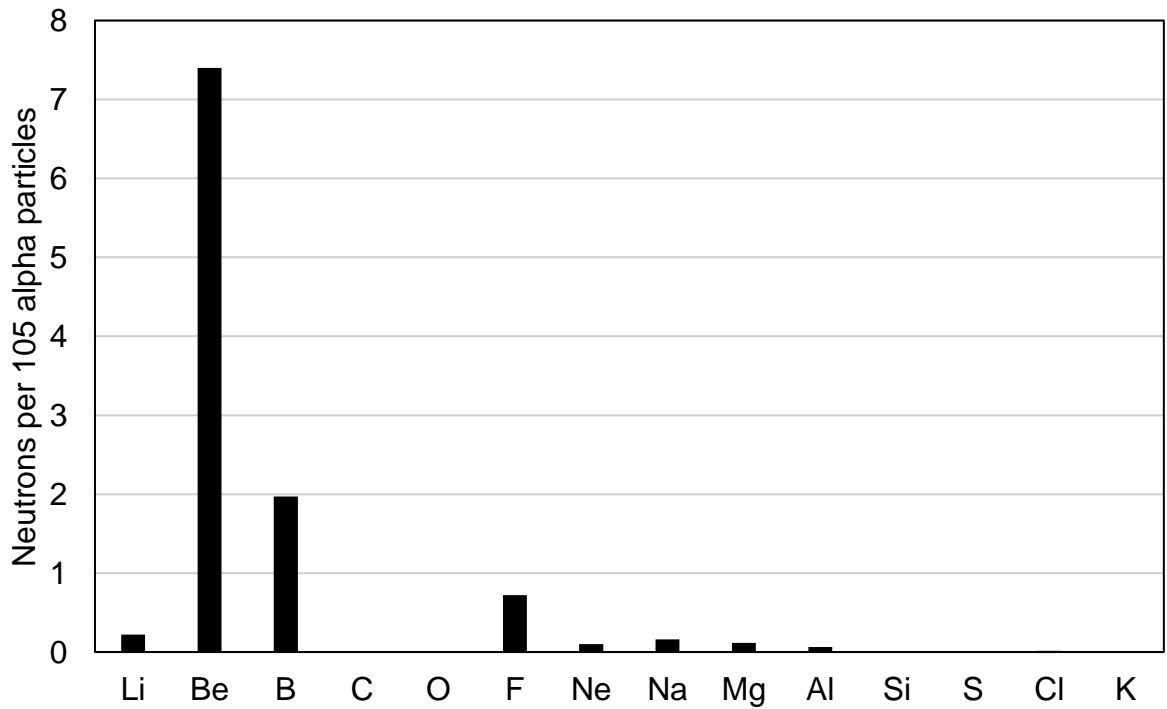

*Figure 2: ($\alpha$,n) yields of light elements at 5.41 MeV*

Atoms of bismuth-209 and beryllium-9 are also able to undergo neutron multiplication reactions such as the (n,2n) reaction to sustain the thermal neutron population in the reaction mixture [6, 40, 41]. This also has the added advantage of lowering the average neutron energy, as the neutron multiplication cross section is much larger for neutrons with energies in the MeV range. This presents a non-competing reaction as the thermal neutrons are able to be captured by bismuth atoms due to the larger neutron capture cross section at lower energies, while fast neutrons are able to be multiplied by bismuth and beryllium atoms due to the larger neutron multiplication cross section at higher energies [15]. Moderation was accomplished by using a counterion with a high hydrogen density in order to saturate the salt with light atoms to slow the neutrons to thermal velocities. Furthermore, beryllium has also been shown to be an excellent neutron moderator for neutrons with higher energies, hence being an added component of moderation within the bismuth salt, serving dual properties of multiplication and moderation in addition to alpha-neutron conversion.

Acetate was used as a counterion in the bismuth salt due to its high hydrogen density for effective moderation, being composed of 5% hydrogen by mass. The hydroxide ion was considered as it has a higher hydrogen density of 15%, but the chemical instability of bismuth beryllium hydroxide made it unsuitable for use in sustaining the chain reaction. Hence, **bismuth beryllium acetate** was chosen as the salt to sustain the chain reaction due to its ability to moderate and multiply neutrons while also possessing a high bismuth percentage and a high alpha-neutron conversion yield.

Bismuth beryllium acetate was prepared by precipitating the acetate salts of bismuth and beryllium respectively from a solution of hot peroxyacetic acid (Fig. 3) in order to minimise the formation of beryllium aerosols. The salt was obtained in good yield (89.3%) as an off-yellow powder.

(Refer to Appendix 2 for detailed preparation procedure)

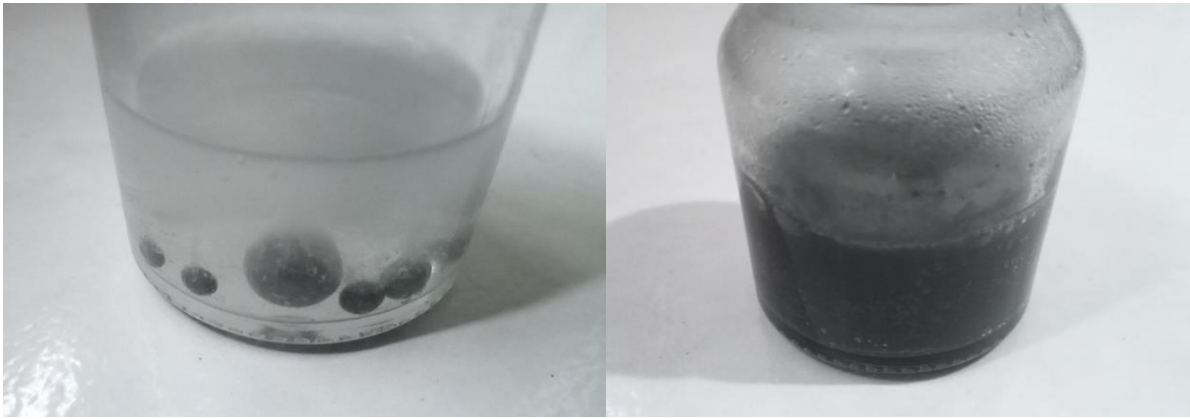

*Figure 3: Dissolution of beryllium metal in peroxyacetic acid (left) and precipitation of bismuth beryllium acetate (right)*

## 2.2. Materials

Radioactive americium-241 mast-mounted and plated sources were obtained from Radwell International MRO. The mast-mounted source was measured to have an activity of 60 μCi, and the two plated sources were measured to have an activity of 10 μCi each. Beryllium foil with a thickness of 0.3mm was purchased from CTPT Inc. Bismuth(III) oxide (>99.9% purity) was purchased from Inoxia Ltd. and used as received without further purification. Beryllium metal (granules) was purchased from Luciteria Inc. and used as received without further purification.

## 2.3. Neutron irradiation setup

Mast-mounted and plated americium-241 sources containing 60 μCi and 10 μCi of $^{241}AmO_2$ respectively plated onto gold foil on a steel base were used as sources of alpha particles for neutron generation, so as to ensure directed alpha particle emission. The neutron source was constructed by exposing a 0.3mm-thick layer of beryllium metal to the alpha radiation from the americium-241 source, allowing fast neutrons to be produced with a mean neutron energy of 4.2 MeV [8]. A 6 cm-thick layer of microcrystalline paraffin wax was used as a neutron moderator to reduce the fast neutrons to thermal energies of < 1 eV to increase the chances of neutron capture by the bismuth-209 atoms [9], and the thermal neutrons produced were used to irradiate a 10 g sample of bismuth beryllium acetate as shown in the irradiation setup diagram below (Fig. 4).

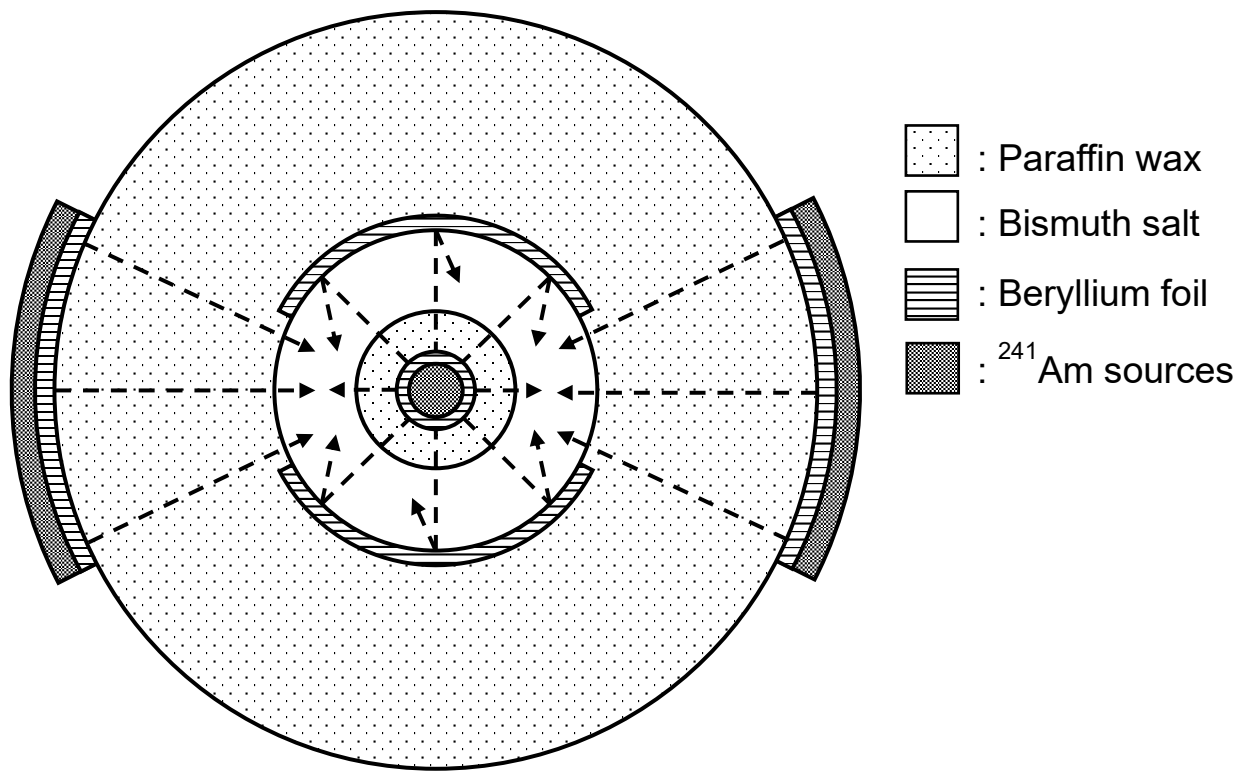

*Figure 4: Neutron irradiation setup*

A central mast mounted neutron source with an activity of 60 microcuries is used to irradiate the surrounding bismuth salt radially, and two other plated neutron source with activities of 10 microcuries each irradiating the bismuth salt from outside. Two beryllium reflectors are also placed on the exposed sides of the bismuth salt in order to shield the excess neutron radiation and to increase neutron flux in the bismuth salt. This setup was adopted to maximise neutron capture in the bismuth salt due to the high neutron opacity of the salt itself which would have cause the outer layers to receive very little neutrons.

2.4. Radiation monitoring and analysis

The potentially harmful neutron, alpha, and beta radiation emitted by the reaction mixture were appropriately shielded as per the guidelines highlighted by the US Nuclear Regulatory Commission [10], the International Atomic Energy Agency [11], and the Singapore Radiation Protection Regulations [12]. The entire irradiation setup was housed in a 15 mm-thick polypropylene container containing a 40 mm-thick paraffin wax layer for radiation shielding.

To qualitatively demonstrate that the trend of the increase in radiation levels is characteristic of the proposed chain reaction, the live radiation levels were monitored with a Geiger-Müller counter and compared to the experimental patterns of growth with an expected rate of increase. The alpha radiation levels were taken as directly proportional to the amount of polonium-210 accumulated in the reaction mixture as the sole alpha-emitting radionuclide in the reaction scheme (excluding the initiator neutron source) is polonium-210. The rate of growth of polonium-210 is derived as the gradient of the polonium accumulation graph.

In order to confirm that polonium-210 has indeed been formed, two methods of characterisation were employed in this study – alpha and beta spectroscopy. Firstly, the most direct method of confirming the presence of polonium-210 was to detect its characteristic alpha peak at 5.30 MeV. This was accomplished via the proximity-count method, where the range of alpha particles in air was correlated to their energy by the equation shown below [42]:

$$R = \frac{0.543E - 0.168}{\rho}$$

where R is the range of the alpha particle, E is its energy, and ρ is the density of the medium. This thus presents a graph (Figure 5) that can be used to map their detected range to their energies, thus giving us the alpha spectrum of the reaction after correcting for the divergence of the radiation with the inverse square law.

Beta spectroscopy was used to identify the parent nuclide of polonium-210, which is bismuth-210 with a maximum beta energy of 1.16 MeV. The stopping power of the beta particles emitted can be calculated using the Bethe-Bloch formula, which correlates to the graph shown below (Figure 6) of the range of beta particles in paraffin wax. As beta particles are emitted over a continuous spectrum unlike alpha particles, there wouldn't be any peak in the beta spectrum of bismuth-210, instead being identified with a drop-off point that correlates to its maximum beta energy at 1.16 MeV.

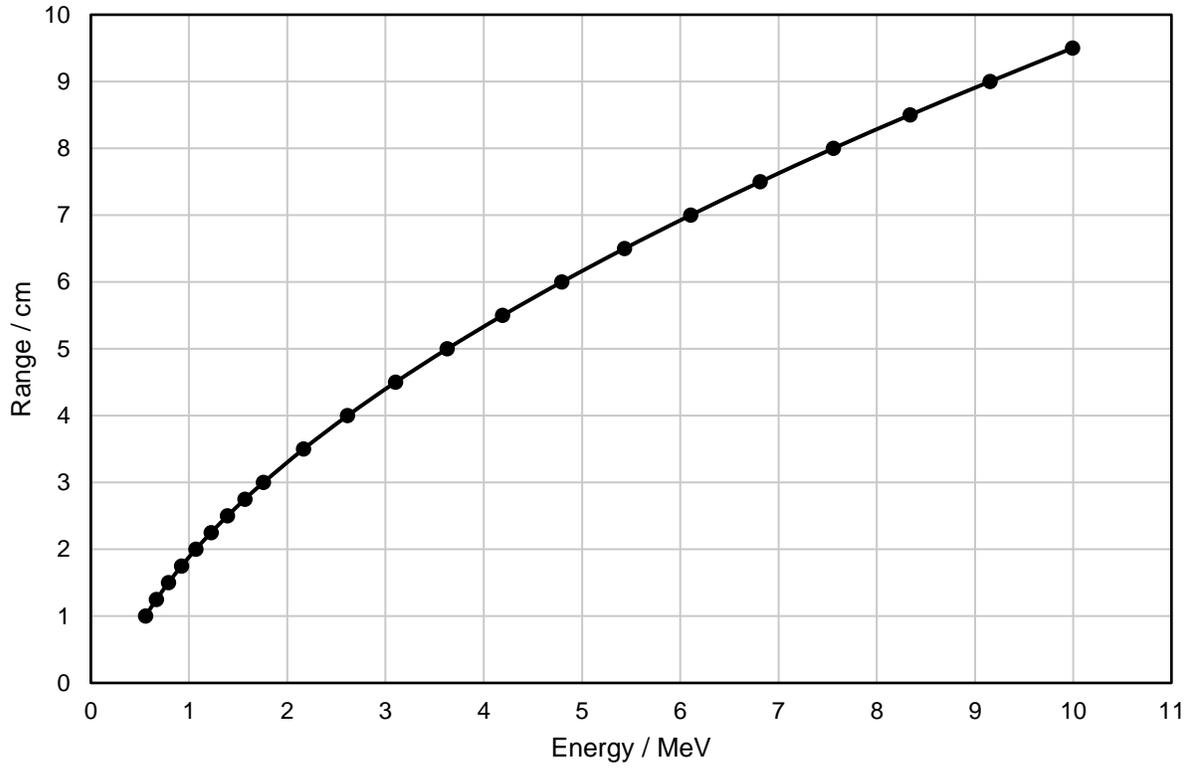

*Figure 5: Range of α particles in air*

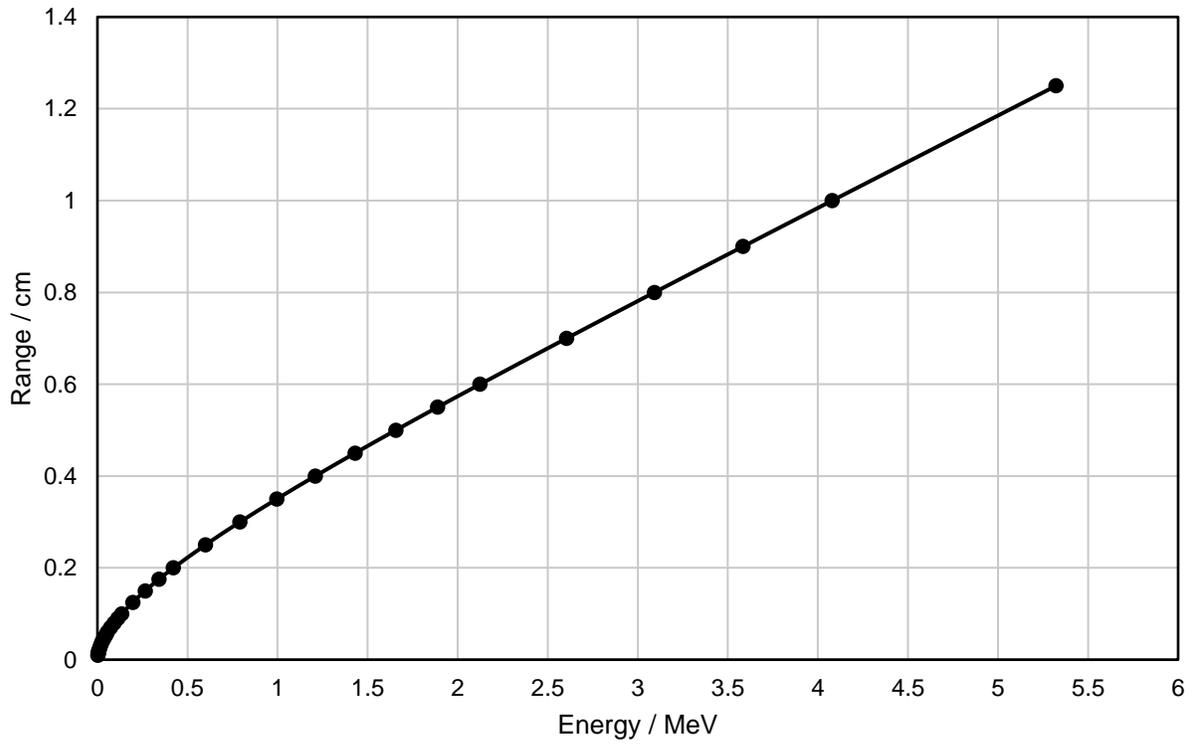

*Figure 6: Range of β- particles in paraffin wax*

## 2.5. Mathematical modelling

In order to quantitatively verify that the chain reaction has indeed occurred, a mathematical model was devised to predict the chain reaction's growth as shown here. Due to the autocatalytic nature of the reaction, the growth of polonium-210 in the reaction mixture is predicted to possess a sigmoidal nature. Its growth mainly depends on 3 factors – the maximum amount of polonium-210 denoted by $L$, and two variables $\alpha$ and $\kappa$ which correspond to neutron population growth and the decay ratio of polonium-210 from bismuth-210 respectively. It should also be noted that an upper limit to the amount of polonium is set by the term $L$ in the chain reaction, as the formation of polonium-210 reaches a maximum before the neutron leakage and decay of initially formed polonium-210 causes the net amount of polonium-210 to be balanced by subsequent generations of the chain reaction which forms more polonium-210.

(For detailed derivation and formulation, see Appendix 1.)

$$N_2 = L \cdot \frac{\alpha e^{\kappa t}}{1 + \alpha e^{\kappa t}} = \frac{L}{1 + \alpha e^{-\kappa t}}$$

$$L = N_{1(0)} \cdot \delta$$

$$\alpha = \frac{\phi \sigma \rho A}{M \varepsilon}$$

$$\kappa = \frac{\lambda_1 n}{\lambda_2}$$

Where:

**$N_2$** is the number of $^{210}$Po atoms at time $t$

**$N_{1(0)}$** is the number of $^{209}$Bi atoms present initially

**δ** is the propagation coefficient

**$\phi$** is the neutron flux of the initial neutron source

**σ** is the neutron capture cross section of $^{209}$Bi

**ρ** is the density of $^{209}$Bi

**A** is Avogadro's number

**M** is the molar mass of $^{209}$Bi

**ε** is the neutron multiplication factor

**$\lambda_1$** is the decay constant of $^{210}$Bi

**$\lambda_2$** is the decay constant of $^{210}$Po

**$n$** is the neutron conversion coefficient

## 3. Results and Discussion

### 3.1. Alpha source characterisation

The alpha spectrum of the mast-mounted and plated americium-241 sources are shown below in Fig. 7 and 8 respectively. The alpha spectrum of the mast-mounted americium-241 source corresponds well to reference literature [17], possessing a characteristic peak at 5.48 MeV. This is indicative of high-purity and thin americium-241 foil being used in the mast-mounted source as a method of ensuring that uniform ionisation of the surrounding medium can be achieved in a small area. Furthermore, the intensity of the alpha spectra obtained from the mast-mounted and plated sources also correspond to the relative activities of the sources (60 and 10 μCi respectively) after factoring in the solid angle of the sources from the detection point. However, the alpha spectrum of the plated americium-241 sources (Fig. 8) showed that peak-splitting was caused by the protective passivation layer on the americium-241 layer on the plates, which were initially designed to ensure that the americium-241 would remain plated to the steel backing. This causes two peaks to be observed in the spectrum at 4.93 and 6.15 MeV respectively.

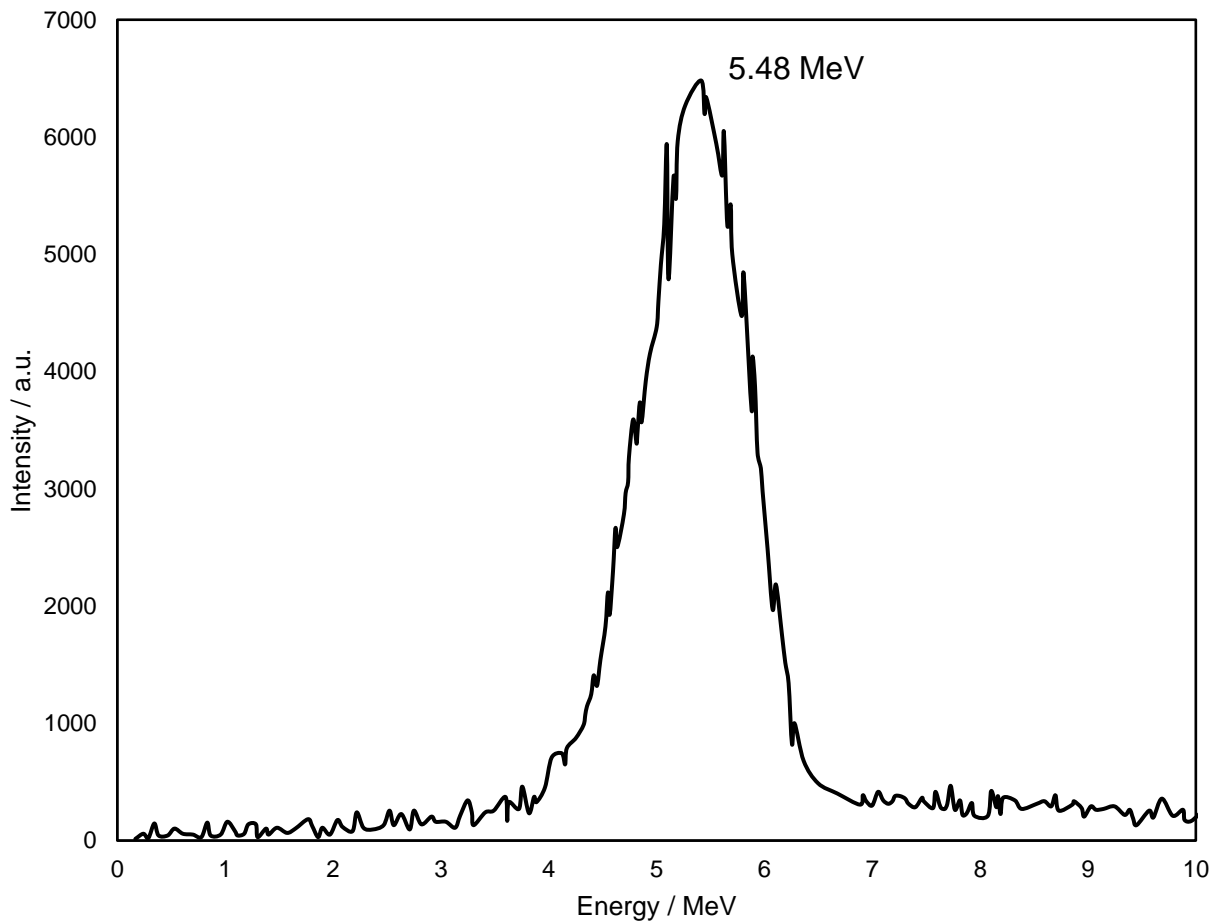

*Figure 7: Alpha spectrum of mast-mounted americium-241 source*

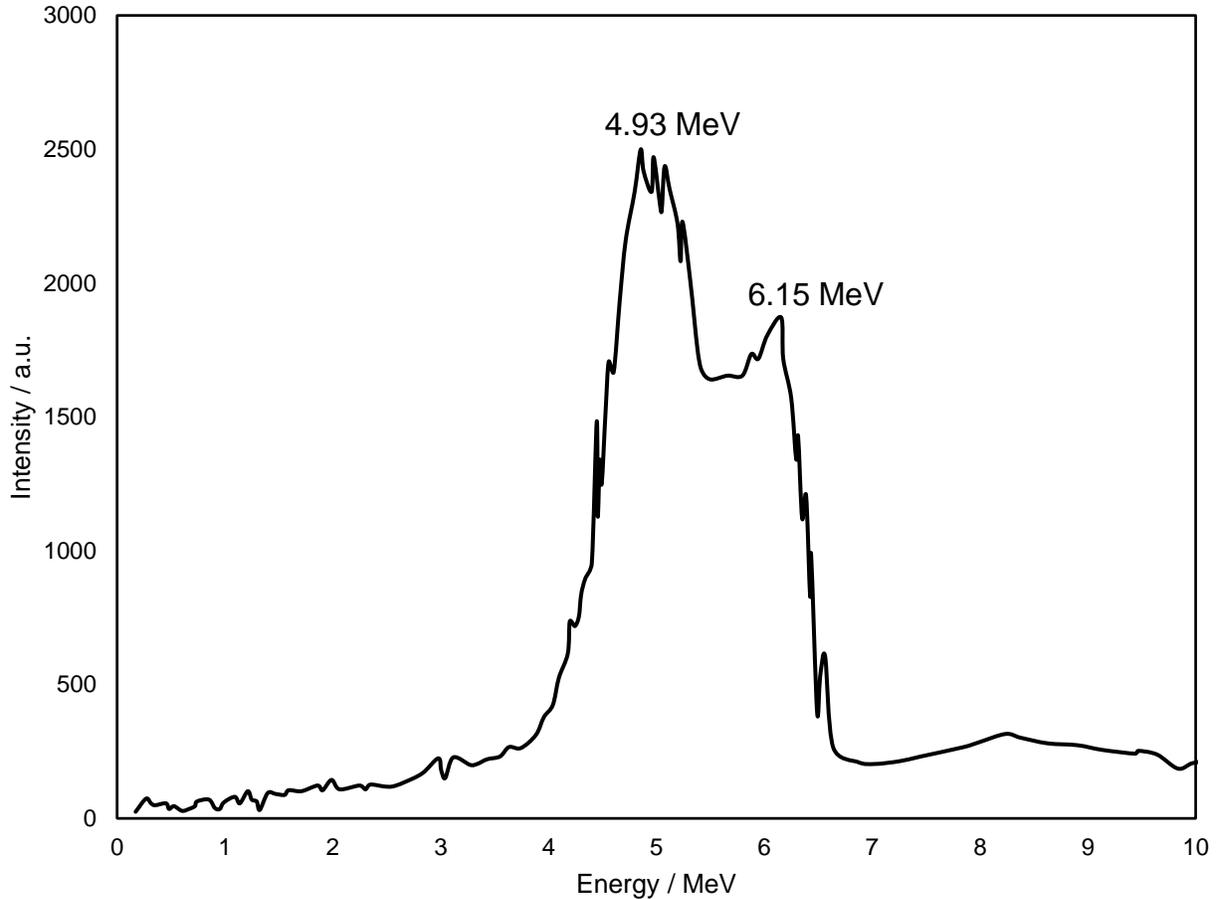

*Figure 8: Alpha spectrum of plated americium-241 source*

3.2. Neutron source characterisation

Utilising the elastic scattering of protons from paraffin wax under neutron irradiation, the fast neutron spectrum of the AmBe neutron source was characterised by mapping the proton recoil energy of the emitted radiation from the moderated neutron source to the corresponding neutron energy in the medium [19]. This yields the neutron spectrum of the emitted neutrons, which was compared to the ISO-8529 reference spectrum for AmBe sources. As shown in Fig. 9, the neutron spectrum of the AmBe source is in good agreement with the reference neutron spectrum [18], thus being suitable for use in initiating the chain reaction. Furthermore, the paraffin wax moderator also served as a neutron shield to prevent fast neutrons from the central mast-mounted neutron source from escaping the setup.

Accounting for the geometric layout of the irradiation setup [22], the thermal neutron flux in the bismuth salt was found to be $2.37 \cdot 10^2$ n cm$^{-2}$ s$^{-1}$. The low neutron flux serves a dual purpose – firstly, it prevents excessive interference with the activity of the polonium-210 produced from the chain reaction so as to ensure that the accumulated polonium-210 can be accurately tracked and measured due to the small neutron capture cross section of bismuth-209 which would cause the amount of polonium-210 formed to be highly sensitive. Secondly, it also demonstrates that the chain reaction is able to be initiated even with weak radioisotopic neutron sources, such that high neutron fluxes are no longer required to produce polonium-210, as the chain reaction is able to propagate on its own through the reaction medium, causing the neutron population to increase in a self-sustaining manner. Once the neutron sources had been constructed and assembled with the irradiation setup, the chain reaction itself was initiated.

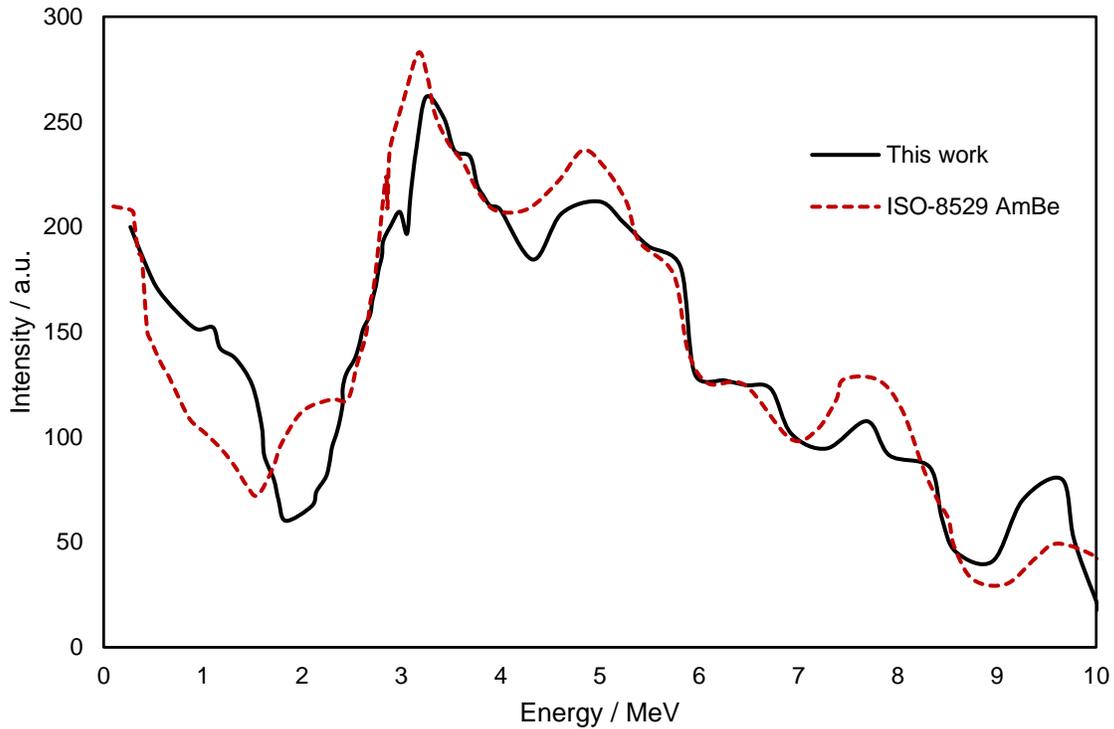

*Figure 9: Neutron spectrum of AmBe source*

3.3. Chain reaction propagation

The characterisation of the chain reaction itself is shown in Fig. 10. As shown, the chain reaction does indeed proceed via a sigmoidal nature in the reaction mixture, starting with an initially low activity due to the small neutron capture cross section of bismuth-209 which results in relatively little polonium-210 being formed. This is analogous to the lag period in a conventional chemical autocatalytic reaction. After some time, the polonium concentration increases exponentially as the chain reaction propagates through the reaction mixture, that is the bismuth beryllium acetate, causing the thermal neutron population to increase exponentially. During this phase, as the atoms of bismuth-209 capture neutrons to form bismuth-210, this does not result in a direct emission of alpha particles but rather a decay according to a first-order mechanism to form polonium-210 which then emits alpha particles, hence causing the exponential phase to be longer than a direct neutron activation reaction due to the consumption and production of the bismuth-210 intermediate.

Furthermore, the decay product of polonium-210 which is lead-206 is also able to cause neutron multiplication via the (n,2n) reaction, further accelerating the reaction as more and more polonium atoms decay. Finally, the reaction slows down after about a week as the limit of polonium accumulation is reached. Not all the bismuth would be converted to polonium at the end of the reaction due to leakage of neutrons out of the reaction mixture coupled with the larger elastic scattering cross section of bismuth-209 compared to its small neutron capture cross section which means that some of the neutrons would simply scatter instead of being absorbed by bismuth-209 atoms [20]. This is represented in the reaction equation as the propagation coefficient, which is proportional to the limit of the amount of polonium-210 that can be extracted from the bismuth-209 feedstock material, which is similar to the conversion ratio in nuclear breeder reactors describing the rate of production of plutonium-239 while accounting for its simultaneous consumption [21]. In a similar pattern, the limit of the amount of polonium-210 able to be produced will also never be equal to the amount of feedstock material due to the decay of the polonium-210 already produced as the chain reaction relies on the alpha particles produced from their decay to be propagated.

Hence, it can be seen from the measurements of the reaction that the chain reaction is indeed being propagated through the bismuth beryllium acetate medium, as evidenced by the rapid sigmoidal increase in activity of the reaction mixture which would otherwise not have occurred if neutron capture by bismuth-209 was the only process taking place.

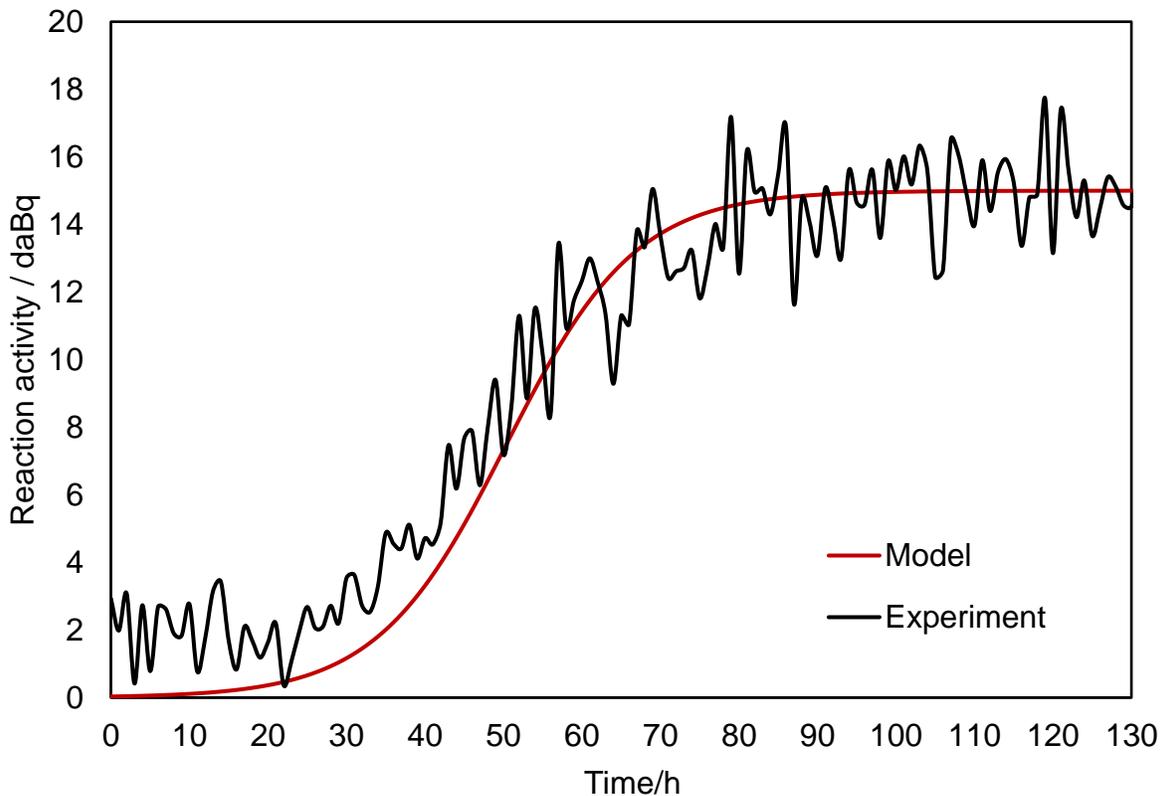

*Figure 10: Theoretical vs experimental results.*

The rate of growth of polonium-210 can also be derived from the gradient of the polonium accumulation curve, shown in Fig. 11 which shows that it reaches a maximum after 50 hours when the chain reaction is still propagating through the reaction mixture while not being impeded by the limits of the medium. In practical applications, this would be the point at which it is the most economical to extract the polonium from the rest of the bismuth salt, as after this point a significant portion of the polonium would have decayed into lead-206 [45, 46]. Furthermore, the rate of growth of polonium-210 in the reaction mixture would also influence the neutron multiplication rate as lead-206 nuclei also possess a relatively large (n,2n) cross section at fast neutron energies. This would result in a positive feedback loop where the increased neutron population causes more nuclei of bismuth-209 to be converted to those of polonium-210, resulting in more lead-206 to be formed upon their decay and further increasing neutron multiplication rates by lead-206. However, this is not observed during normal neutron irradiation of bismuth-209 due to the balance by decreasing the amount of bismuth-209 available to multiply neutrons.

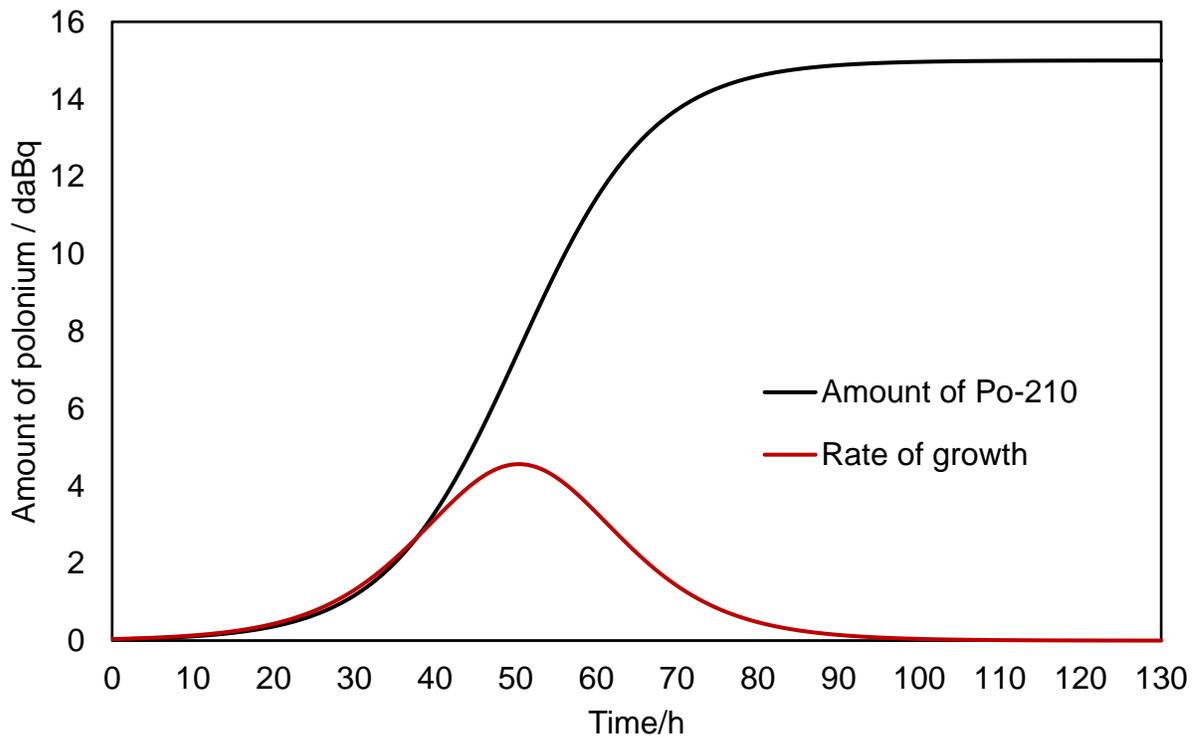

*Figure 11: Rate of growth of polonium-210*

The rate of growth of polonium-210 is also constrained by several factors, most importantly the variable κ which is affected by the decay constants of polonium-210 and bismuth-210, as well as the neutron conversion coefficient. As the decay constants of both nuclides remain unchanged, the neutron conversion coefficient, i.e. the fraction of neutrons from a generation of the chain reaction that get captured by bismuth atoms from subsequent generations, would be the main influencing factor in determining how fast polonium-210 accumulates up to the limit *L*.

The polonium accumulation curve of the chain reaction was also compared to that of a normal irradiation scenario shown in Fig. 12, which shows that conventional irradiation methods allow for a greater initial rate of growth of polonium due to the larger number of bismuth atoms exposed to the neutron radiation closer to the neutron source without the interference of other atoms [43, 44], which is not the case in the chain reaction due to some neutrons being absorbed by auxiliary atoms, e.g. hydrogen and beryllium atoms. However, the chain reaction quickly surpasses it because of the rapid and exponential propagation of the polonium conversion reaction throughout the reaction medium due to the polonium-210 atoms formed in one generation acting as the "source" for the next, thus causing a cascade reaction in the medium which is faster than the capture of individual neutrons by bismuth-209 atoms.

Furthermore, the production of polonium-210 is also influenced by the penetration of neutrons through the reaction medium, e.g. bismuth metal or salt, and thus the chain reaction would also produce polonium-210 at a faster rate as the neutrons produced in the chain reaction would be able to saturate the reaction mixture faster as compared to the mixture being irradiated by a point source [47, 48]. Hence, this chain reaction would allow for more polonium-210 to be produced at a faster rate after an initial lag period as compared to a normal irradiation route, which would allow dependence on high-flux nuclear reactors for the production of polonium-210 to be reduced as it is able to be produced using weak radioisotopic neutron sources.

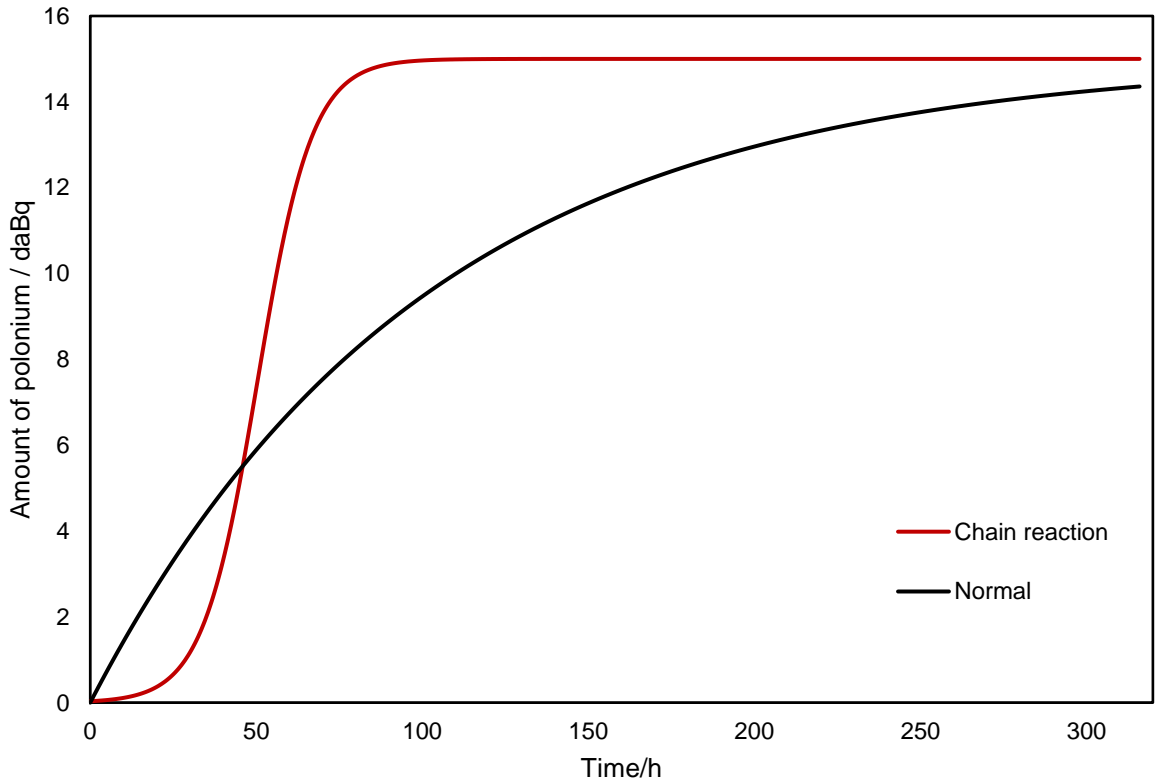

*Figure 12: Growth of polonium-210 under chain reaction and normal conditions*

## 3.4. Alpha spectrum ($^{210}$Po)

As an additional method of ascertaining that the chain reaction had indeed occurred, the presence of polonium-210 was confirmed in the reaction mixture using alpha spectroscopy. As seen in the alpha spectrum of the reaction mixture given in Fig. 13, the distinct peak at 5.30 MeV is indicative that polonium-210 had been produced by the chain reaction [24], as the only alpha emitter involved in the mechanism of the reaction (excluding the initiator source) is polonium-210. The alpha emissions from polonium-210 were differentiated from the beta emissions of bismuth-210 by correlating the respective activities of the reaction mixture at consistently varied distances from the source, thereby allowing two separate graphs to be obtained.

Additionally, it was also seen that the alpha peak of the reaction mixture was considerably broader than expected due to the slowing down of alpha particles from scattering effects of the beryllium and hydrogen atoms [49, 50], which would have caused a large range of energies to be detected. Hence, purification of the polonium-210 produced from the reaction would aid in more accurate alpha spectroscopy measurements.

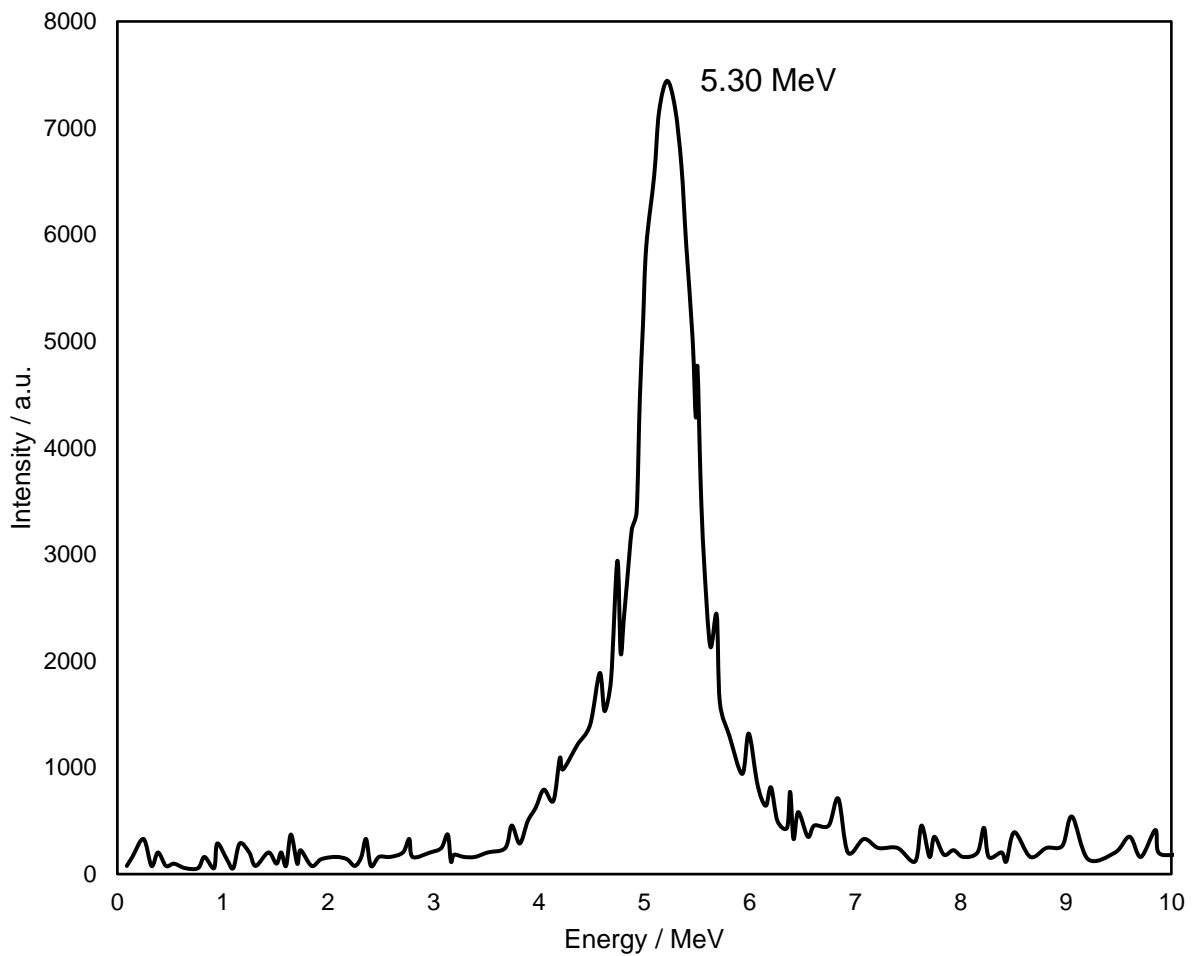

*Figure 13: Alpha spectrum of reaction mixture*

## 3.5. Beta spectrum ($^{210}$Bi)

The beta spectrum of the reaction mixture is also shown below. The beta spectrum graph in Fig. 14 shows a maximum energy of 1.16 MeV, which corresponds to the maximum beta energy of bismuth-210 [25] i.e. the parent nuclide of polonium-210, thus showing that the reaction was indeed producing bismuth-210 from the neutron capture of bismuth-209. However, this does not rule out the presence of other beta-emitting neutron activation products such as carbon-13 and hydrogen-3 with lower maximum beta energies (0.16 and 0.018 MeV respectively). Nevertheless, this also serves as an additional method of confirming that bismuth-210 is indeed present in the reaction mixture, which would indicate that polonium-210 is being produced. Furthermore, the beta spectrum also indicates that the chain reaction is progressing due to the weaker beta spectrum obtained during the stabilisation period of the reaction, which was attributed to the decreasing amounts of bismuth-210 present due to its relatively short half-life compared to polonium-210, as well as a significant portion of the bismuth-210 having decayed to polonium-210.

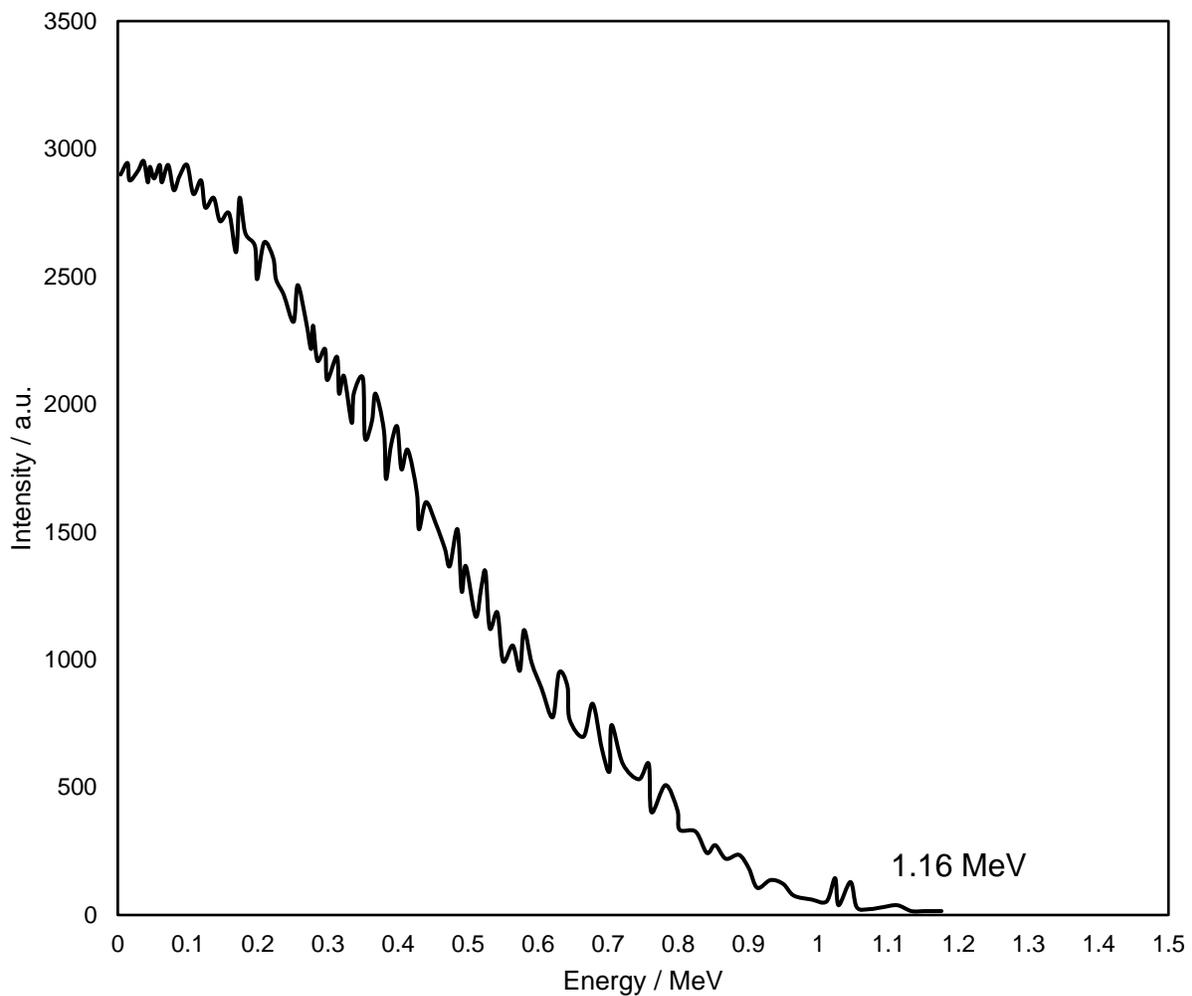

*Figure 14: Beta spectrum of reaction mixture*

## 4. Conclusions

Self-sustaining neutron multiplication has thus been proven to be achieved in a medium of bismuth beryllium acetate, producing the industrially and politically significant radionuclide polonium-210 in substantial quantities as a byproduct of the reaction. This chain reaction, which proceeds via a mechanism involving alternating (α, n) and alpha decay reactions, was confirmed and validated via the monitoring of the activity levels of the reaction mixture, as well as via characterisation of the alpha particle energies emitted from the mixture that were correlated to the alpha emission spectra of polonium-210 and the beta spectrum of the parent nuclide bismuth-210. The activity levels of the reaction mixture were shown to agree with a mathematical model of the chain reaction, notably demonstrating a sigmoidal alpha activity which was ascribed to the increasing quantities of polonium-210 produced.

This previously-unreported reaction also offers a relatively inexpensive and rapid method of producing polonium-210 in appreciable quantities, as highlighted by the alpha-emission characterisation of the reaction. Furthermore, purification of the polonium-210 produced is fairly straightforward and high-yielding, due to the crystalline nature of the reagents involved that would allow for direct reduction of the salt, as opposed to conventional purification methods of polonium that require separation of the raw metal product from bismuth metal feedstock.

## 5. Future Work

Due to the strategic importance of the radionuclide polonium-210, e.g. in the nuclear weapon complex, political operations, satellite power generators, radiotherapy etc, this chain reaction would have many industrial applications. For example, the scalability of the reaction could be further researched in order to produce polonium-210 on an industrial level, which would alleviate the current scarcity of polonium as only 100 grams of it are produced annually [14]. This reaction could also be exploited to construct more efficient radioisotopic thermoelectric generators (RTGs) for use on board satellites [23], as the short half-life of polonium-210 ($T_{1/2}$ = 138 days) which hinders its incorporation into RTGs for long journeys would be resolved by the continuous generation of polonium from the bismuth salt feedstock, allowing the polonium to be replenished.